\documentclass[twocolumn,showpacs,preprintnumbers,amsmath,amssymb,prb]{revtex4}
\usepackage{graphicx}% Include figure files
\usepackage{dcolumn}% Align table columns on decimal point
\usepackage{bm}% bold math

\newcommand{\nil}{\hspace*{0em}}
\begin{document}
\title{Covalency effects on the magnetism of EuRh$_{\mathbf{2}}$P$_{\!\mathbf{2}}$}
\author{Robert Schmitz and Erwin M\"uller-Hartmann}
\affiliation{Institut f\"ur Theoretische Physik,
Universit\"at zu K\"oln, Z\"ulpicher Stra{\ss}e 77, 50937 K\"oln,
Germany}

\date{\today}

\begin{abstract}
In experiments, the ternary Eu pnictide EuRh$_2$P$_{\!2}$
shows an unusual coexistence of a non-integral Eu valence
of about 2.2 and a rather high N$\acute{\rm{e}}$el temperature of 50 K.
In this paper, we present a model which explains the non-integral Eu
valence via covalent bonding of the Eu 4f-orbitals to P$_{\!2}$
molecular orbitals. In contrast to intermediate valence models
where the hybridization with delocalized conduction band electrons
is known to suppress magnetic ordering temperatures to at most a
few Kelvin, covalent hybridization to the localized P$_{\!2}$
orbitals avoids this suppression. Using perturbation theory we
calculate the valence, the high temperature susceptibility, the Eu
single-ion anisotropy and the superexchange couplings of nearest
and next-nearest neighbouring Eu ions. The model predicts a
tetragonal anisotropy of the Curie constants. We suggest an
experimental investigation of this anisotropy using single
crystals. From experimental values of the valence and the two
Curie constants, the three free parameters of our model can be
determined.
\end{abstract}


\maketitle

\section{Introduction}

\subsection*{Non-integral valence, intermediate valence, and magnetic order}

A general question of interest is how a non-integral valence of localized ions in a solid influences the possibility of magnetic order. Concerning this subject, an earlier paper reported the anomalous valence state of Eu in EuRh$_2$P$_{\!2}$. \cite{michels} EuRh$_2$P$_{\!2}$ was characterized to show intermediate-valent and probably also covalent properties. A coexistence of a non-integral Eu valence of 2.2 and antiferromagnetic order up to 50 K was reported.

Simply given, a non-integral valence means that the mean total occupation number of the ionic electronic levels is non-integral. This might have several physical reasons. In particular, {\it{intermediate valence}} is the hybridization of localized ionic states with the strongly {\it{delocalized}} conduction band. It is experimentally evident and theoretically well-understood that there is a strong competition between intermediate valence and magnetic order. Intermediate valence is known to suppress magnetic ordering temperatures typically to at most a {\it{few}} K. As an experimental example, TmSe is intermediate-valent and has a N$\acute{\rm{e}}$el temperature below 3.5 K. \cite{holmo}

A detailed theoretical approach to intermediate valence is the extended s-f model. \cite{nolting2} Using this model a similarity to the Kondo effect is shown: grossly spoken, information about the electron spin is washed out by the delocalization of the conduction band states. The extended s-f model allows one to take a ferromagnetic exchange between ionic f and conducting s states into account, which is typical for intermediate-valent Eu ions. Hence, it is more realistic than an Anderson model, which always implies an antiferromagnetic s-f coupling. \cite{schriewo}

In agreement with experiment, the extended s-f model also explains why the magnetic ordering temperature of an intermediate-valent system is enhanced drastically by mechanical or chemical high pressure. In addition, a given intermediate valence is pressure-sensitive itself because the energy levels of the localized states can be pressed towards the conduction band. These aspects are not too important for the present paper because we will concentrate neither on intermediate valence nor on anomalous pressure.

If a non-integral valence is caused by {\it{covalence, no strongly delocalized states}} are involved in the underlying hybridization. Hence, a characteristically different approach will be required to understand such systems.

The interpretation of EuRh$_2$P$_{\!2}$ is insufficient so far. On
the one hand, if EuRh$_2$P$_{\!2}$ was a system in which a Eu
intermediate valence of 2.2 and magnetic order coexist up to 50 K
without any restrictions, this would mean a small sensation. Not
only was such a behaviour experimentally unusual but also
theoretically not understood, because in this case the ordering
temperature was an order of magnitude higher than for typically
realistic parameter sets of the extended s-f model.
\cite{nolting2} Even for extreme choices of the model parameters
(in case of a Eu intermediate valence of 2.2), an upper boundary
of about 15 K for the ordering temperature is estimated. A new
concept of intermediate valence would have to be found to explain
the experiment. On the other hand, no detailed model is available
to investigate the counter-perspective, the influence of the
covalence on the magnetism of EuRh$_2$P$_{\!2}$.

In the following, we will motivate the use of a covalent scenario {\it{instead}} of intermediate valence as the starting-point for the investigation of EuRh$_2$P$_{\!2}$. This is in contrast to former publications \cite{michels,wurth,johrendt,huhnt0} and also refers to unpublished experimental material. \cite{niemoeller,schuette,chefki,huhnt}

\subsection*{The covalence of EuRh$_2$P$_{\!2}$}

We take into account the following experiments on EuRh$_2$P$_{\!2}$: measurements of the magnetic susceptibility, \cite{michels,schuette} of the crystal structure, \cite{wurth,huhnt0,huhnt} L$_{\rm{III}}$ x-ray absorption, \cite{niemoeller} and M\"o{\ss}bauer spectroscopy. \cite{chefki}

\begin{figure}
\includegraphics[width=7cm]{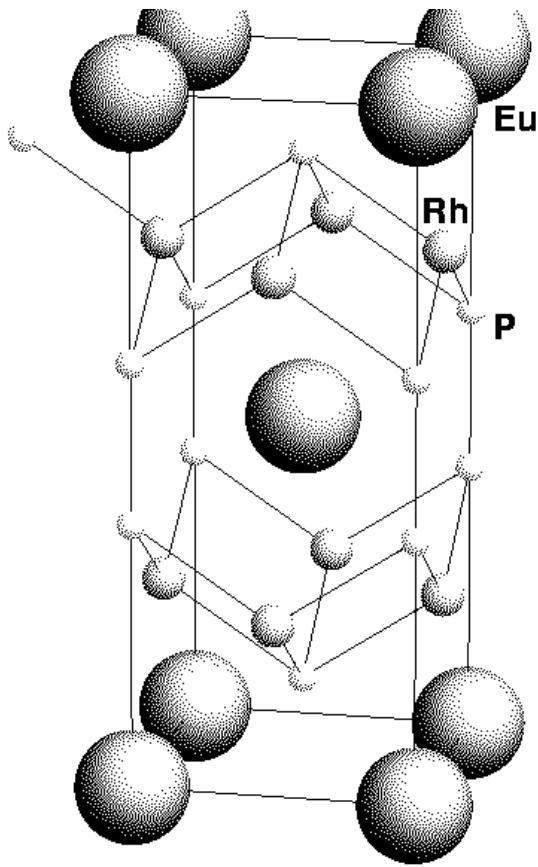}
 \caption{Crystal structure of EuRh$_2$P$_{\!2}$.}\label{crstruc}
\end{figure}

The properties of EuRh$_2$P$_{\!2}$ can be unraveled by studying the influences of temperature and doping, especially with As. Depending on these two parameters, in experiments a structural phase transition of first order is observed whereas the lattice type, a body-centered tetragonal ThCr$_2$Si$_{2}$ structure (figure \ref{crstruc}), does not change.

The structural phases $\alpha$ and $\beta$ differ by a strongly
anisotropic jump of the lattice parameters and by the electronic
bonding conditions. This leads in the $\beta$ phase to a
non-integral Eu valence of 2.2. The lattice cells show a slight
elongation in the square plane but a considerable compression
along the tetragonal axis \cite{wurth,huhnt0,huhnt} compared to
the $\alpha$ phase, which contains divalent Eu ions. In both
phases, the valence is homogeneous, i.\,e., all Eu ions are
equivalent.

In the $\beta$ modification two nearest neighbouring P atoms form a single molecular bond \cite{huhnt0,huhnt}---oriented along the tetragonal axis---which does not appear in the $\alpha$ phase. That corresponds to a charge transfer from the phosphorus to the conduction band, which leaves holes in P$_{\!2}$ molecular states. These are available for occupation by one fluctuating Eu electron each. In this way, in the $\beta$ phase there is a covalent hybridization between the magnetic Eu ions and P$_{\!2}$ molecules. The Rh ions do not participate in the magnetic properties of EuRh$_2$P$_{\!2}$.

Both phases $\alpha$ and $\beta$ exhibit a magnetic phase transition
from para- to antiferromagnetism at a N$\acute{\rm{e}}$el temperature
$T_{\rm{N}}$ (phase diagram: figure 2%\ref{phdiag}
). With decreasing As doping $T_{\rm{N}}$ drops somewhat at the
$\alpha$-$\beta$ phase boundary but does not change its order of
magnitude. According to the extended s-f model this effect of the
valence transition on the ordering temperature is too small to be
consistent with intermediate valence.

The valence measurements on EuRh$_2$P$_{\!2}$, in particular L$_{\rm{III}}$ x-ray absorption and M\"o{\ss}bauer spectroscopy, can not distinguish between a non-integral valence of covalent and intermediate-valent origin. The reduced magnetic moment of the $\beta$ phase compared to the divalent Eu moment of the $\alpha$ phase does not clarify the nature of the non-integral valence either. Furthermore, the valence cannot be derived from the magnetic Eu moment alone because in the non intermediate-valent case a non-trivial contribution by the P$_{\!2}$ states must be considered. This quantity is unknown.

The covalent bonding scenario between Eu and P$_{\!2}$ states persued in the present paper was derived in reference \onlinecite{huhnt} from bonding lengths and is supported by further aspects. In particular, the pressure dependence \cite{chefki} of the valence, which we express as the deviation $\Delta W$ from the divalent state, leads to the conclusion that the intermediate-valent part of the total Eu valence amounts to at most
\begin{equation} \label{MV}
\Delta W |_{\rm{iv}} \approx 0.05.
\end{equation}

This is the precision with which the Eu valence is determined by M\"o{\ss}bauer spectroscopy. \cite{chefki} Reference \onlinecite{chefki} states {\it{no}} intermediate valence within this precision because in M\"o{\ss}bauer spectroscopy the Eu valence does not change over the range between ambient pressure and 5 GPa. This is the expectation in absence of a conduction band hybridization.

As the results of L$_{\rm{III}}$ x-ray absorption \cite{niemoeller} may be influenced by final-state effects, we assume an error bar for the non-integral Eu valence of the $\beta$ phase including intermediate valence as well as covalence and estimate the valence shift as:
\begin{equation} \label{ValDev}
\Delta W=0.20 \pm 0.05.
\end{equation}

We will not carry out an error calculation in closed form but will select several discrete values for parameters of the model, which will be constituted neglecting $\Delta W |_{\rm{iv}}$. Furthermore, we ignore the weak temperature dependence of the valence which is considerably less significant in the $\beta$ phase %(probably of the order 0.01 \cite{niemoeller})
than the precision of 0.05. \cite{chefki}

The mechanism and the geometric structure of {\it{magnetic ordering}} in EuRh$_2$P$_{\!2}$ are almost completely unknown. Experimentally, antiferromagnetism below $T_{\rm{N}}$ has been concluded, as well as the existence of---unspecified---ferromagnetic couplings because of an anomalous paramagnetic Curie temperature. \cite{schuette} In addition, the importance of a phosphorus-mediated superexchange between the Eu ions has been shown qualitatively via the sensitivity of the M\"o{\ss}bauer magnetic hyperfine field (and $T_{\rm{N}}$) to pressure, \cite{chefki} which is not observed in the reference system EuRh$_2$As$_{2}$.

\begin{figure}
\includegraphics[width=7cm]{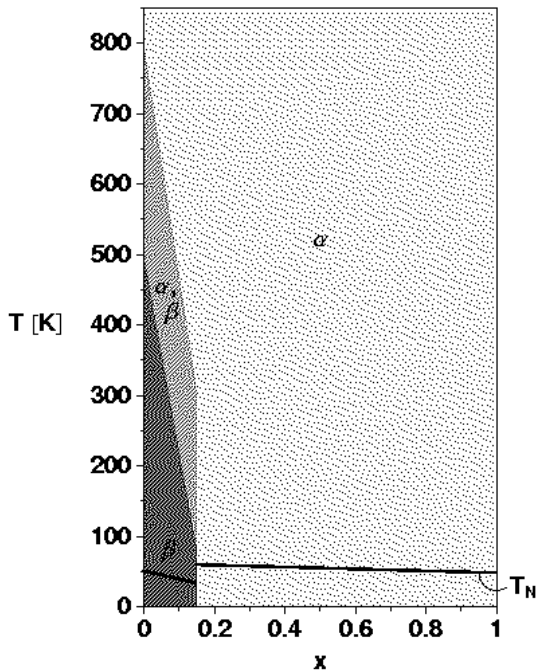}
\caption{Phase diagram of EuRh$_2^{\,} ( $As$_x^{\,}$P$_{1-x}^{\,} ){\nil}_2^{\,}$. \cite{schuette} The mid-grey sector corresponds to a coexistence due to hysteresis.}
\end{figure}

We conclude that the covalence of EuRh$_2$P$_{\!2}$ has to be significant because the intermediate-valent part is small and the reduced magnetic moment, the existence of the superexchange, and the bonding scenario have been shown to be mutually consistent with the covalence.

A possibility other than covalence or intermediate valence which can cause a measured non-integral valence should be excluded in EuRh$_2$P$_{\!2}$: In Eu$\big($Pd$_{0,7}$Au$_{0,3}\big)\nil_{2\,}$Si$_{2\,}$ similar L$_{\rm{III}}$ x-ray absorption and M\"o{\ss}bauer results in spite of a {\it{divalent}} magnetic Eu moment are found at ambient pressure and are explained finally by an anomalous spatial extension of the 4f shell, whereas the system becomes intermediate-valent under high pressure. \cite{abd4} In contrast to EuRh$_2$P$_2$, in Eu$\big($Pd$_{0,7}$Au$_{0,3}\big)\nil_{2\,}$Si$_{2\,}$ no covalent bonding partners are available for the Eu ions.

Under pressure beyond 5 GPa, intermediate valence in EuRh$_2$P$_{\!2}$ acquires, at least qualitatively, the same relevance as the covalence but stays less important than the latter at ambient pressure. \cite{chefki} The corresponding characteristic decrease of $T_{\rm{N}}$ under increasing pressure is understood by the extended s-f model.

Band structure calculations of EuRh$_2$P$_{\!2}$ have only been performed for the integral-valent $\alpha$ (high temperature) phase. \cite{johrendt}

The magnetic ordering of EuRh$_2$P$_{\!2}$ is caused both by super- and
indirect exchange. 
The latter aspect follows because the reference system
EuRh$_2$As$_{2}$ exhibits almost the same $T_{\rm{N}}$ (48\,K) but
neither superexchange \cite{chefki} nor an anomaly of the
paramagnetic Curie temperature. \cite{schuette}

In order to treat the covalence effects of EuRh$_2$P$_{\!2}$
perturbatively from divalent ionic ground states we view the
crystal as built by inactive two-dimensional metallic Rh planes
and by quasi-two-dimensional insulating EuP$_{\!2}$ planes. The
covalent hybridization in these planes is much more important than
the one between different EuP$_{\!2}$ planes. This is concluded
from the Eu-P distances, which in the $\beta$ phase near the phase
boundary are: 3.88 {\AA} between the planes and 3.10 {\AA} in the
plane.

As the starting-point we view the properties of an arbitrary
single Eu ion in interaction with the four neighbouring P$_{\!2}$
molecular ions which form a tetragonal cage around it. This
interaction describes the high-temperature paramagnetism of the
crystal to leading order. We begin with three unknown model
parameters and reduce their number finally to one after
calculating two quantities for which experimental values are
available: the valence and the paramagnetic susceptibility (as an
avarage over the three spatial directions). We also calculate the
single-ion anisotropies of the Eu ions, which are experimentally
unknown.

Using the same model parameters, since thermal expansion is very small in the temperature range between 0 and about 400 K, \cite{huhnt0} we calculate the superexchange parameters between nearest and next-nearest neighbouring Eu ions.

\section{Model}

The $\alpha$ phase of EuRh$_2$P$_{\!2}$ has got the formal valences \cite{huhnt}:
\begin{eqnarray}
\mbox{Eu}^{2+}\big(\mbox{Rh}^{2+}\big)_2\hspace*{-1ex}\underbrace{\big(\mbox{P}^{3-}\big)_2}_{\mbox{\footnotesize{full 3p shell}}}\hspace*{-1ex}.
\end{eqnarray}
This corresponds to the divalent ground state of a Eu ion:
\begin{eqnarray}
4{\rm{f}}^7,\; S=\mbox{$\frac{7}{2}$},\;L=0,\;J=S.
\end{eqnarray}
In the covalent $\beta$ phase there is a fluctuation between the formal valences: \cite{huhnt}
\begin{eqnarray}
\mbox{Eu}^{2+}\big(\mbox{Rh}^{+}\big)_2\!\underbrace{\big(\mbox{P}_2\big)^{4-}}_{\mbox{\footnotesize1 hole per} \atop \mbox{\footnotesize{P core}}} & &  \quad \longleftrightarrow\;\mbox{ Eu}^{3+}\big(\mbox{Rh}^{+}\big)_2\!\underbrace{\big(\mbox{P}_2\big)^{5-}}_{\mbox{\footnotesize1 hole per} \atop \mbox{\footnotesize{molecule ion}}}\hspace*{-1ex}.
\end{eqnarray}

The hybridization of the Eu ions is described perturbatively---starting from the divalent configuration---via a quantum mechanical admixture of trivalent states:
\begin{eqnarray}
4{\rm{f}}^6,\; S=3,\;L=3,\;J=0...6.
\end{eqnarray}

Hund's rule correlations are taken fully into account. We use a Land$\acute{\rm{e}}$ approximation for the energies $E_J$ of the Eu$^{3+}$ ground state $J=0$ and the low excitations $J>0$:
\begin{equation}
E_J=\Delta E\,[1+XJ(J+1)].
\end{equation}

$E_{J=0}$ is shifted by an unknown charge transfer energy $\Delta E$ with respect to the divalent ground state, which is the zero point in our calculation ($E_{2+}=0$). The intraionic spin-orbit part of the levels $E_J$ is given by
\begin{equation}
X=\frac{\xi}{42 \Delta E}, \quad \xi=7960\,\mbox{K}
\end{equation}
which was fitted to optical measurements \cite{chang} by a least
squares fit. $\xi$ is the spin-orbit coupling parameter
($H_{\rm{so}}=\xi {\mathbf{L}}{\mathbf{S}}$ is the Hamiltonian of
the spin-orbit coupling for a single Eu ion).

After an electron hops from the Eu ion to a phosphorus ion, it occupies an anti-bonding molecular orbital (MO) state which is considered as a linear combination of two 3p$_z$ atomic orbitals due to the MO method and which is odd under a reflection with respect to the $a^2$ ($xy$) plane. The bonding 3p$_z$ MOs are always filled.

As the perturbation term which expresses the covalent hybridization we introduce a hopping operator between a Eu ion and four neighbouring P$_{\!2}$ molecular ions:
\begin{equation}
V=\hspace*{-1.3ex}\sum_{k=1,2,3,4 \atop
\sigma=\uparrow,\,\downarrow}\hspace*{-1.3ex}
\big[t_0^{\,}f^\dag_{0\sigma}+(-1)^k\,t_2^{\,}\big(f^\dag_{-2\sigma}
+f^\dag_{2\sigma}\big)\big]p^{(k)}_\sigma+\mbox{H.\,c.}
\label{perturbation}
\end{equation}

$p$ and $f$ denote the annihilators corresponding to the single-particle states. Because of symmetries (time reversal of the crystal Hamiltonian, reflection with respect to $xy$ and $xz$ plane) contributions due to hopping amplitudes $t_m$ for magnetic quantum numbers $m \neq 0, \pm 2$ are excluded and we have $t_{-2}=t_2$. {\small$p^{(k)}_\sigma$} relates to the antibonding MO of the $k$th neighbour. There are three unknown model parameters: $X$, $t_0$, and $t_2$. The hopping amplitudes can be chosen real.

We calculate the matrix elements of effective operators via standard perturbation theory for a degenerate system and to leading order in the perturbation. We use a formulation due to Takahashi \cite{takahashi} involving a unitary transformation $\Gamma$ which maps the perturbed problem onto the unperturbed ground state space. $\Gamma$ is given as a power series in terms of the unperturbed Hamiltonian---corresponding here to the energies $E_{2+}=0$ and $E_J$---and the perturbation operator $V$. Any operator $A$ in the Hilbert space of the perturbed states is treated as an effective operator $a=\Gamma^\dag A\, \Gamma$ in the ground-state space.

In the calculation, the matrix elements of the effective operators are expressed in terms of the matrix elements of the $f$ creators (which appear in the perturbation term $V$) between the correlated many-body states of the Eu$^{2+}$ and Eu$^{3+}$ ions. The matrix elements are evaluated via Clebsch-Gordan coefficients and the Wigner-Eckart theorem. Explicitly, using Clebsch-Gordan coefficients the $f$ creators are transformed from the $m\sigma$ to the $jj_z$ basis:
\begin{equation}
f_{m \sigma}^{\dag}=\hspace*{-0.5em}\sum_{j=\frac{5}{2},\frac{7}{2} \atop j_z=m+\sigma} \hspace*{-0.5em}\big<3m \mbox{$\frac{1}{2}$} \sigma \big| jj_z \big> f_{jj_z}^{\dag}.
\end{equation}
The Wigner-Eckart theorem gives:
\begin{equation}
\big<\mbox{$\frac{7}{2}$} M \big| f_{jj_z}^{\dag} \big| JJ_z \big> = \big \| f \big \|_{jJ} \big<JJ_z jj_z \big|\mbox{$\frac{7}{2}$} M \big>,
\end{equation}
where $\big|\frac{7}{2} M \big>$ is a state of the Eu$^{2+}$ configuration ($M=-7/2...7/2$), $\big| JJ_z \big>$ is a state of the Eu$^{3+}$ configuration ($J=0...6,J_z=-J...J$), and $\big \| f \big \|_{jJ} $ is a reduced matrix element.

\section{Single-ion effects}

We calculate the matrix elements of the effective operators of the valence and the magnetization due to hopping of second order in $V$, which is the leading order of the hybridization for the single-ion effects. For this calculation it is sufficient to apply the unitary operator $\Gamma$ to first order.

We have calculated the effective single-ion Hamiltonian to second order in $V$. According to this calculation, which we do not present in detail, the octets of the unperturbed Eu$^{2+}$ ions split into four Kramers doublets which, however, remain quasi-degenerate, i.\,e. the splitting of these doublets is very small compared to the temperature. Hence, we will use the average energy of the Kramers doublets for thermodynamic averaging.

\subsection*{Effective valence}

The covalent admixture of Eu$^{3+}$ states shifts the valence from 2 to a larger value. In order to calculate the model valence we use the valence operator:
\begin{equation}
W=3-P_0\,.
\end{equation}

$P_0$ is the projector onto the unperturbed ground-state space. \cite{takahashi} In the framework of our perturbation method \cite{takahashi} we use the effective valence operator $w=\Gamma^\dag W\, \Gamma$ to second order in $V$. Thermodynamic averaging due to the mean eigenvalues of the effective Hamiltonian of second order gives the mean valence:
\begin{equation}
\big<W\big>=\mbox{$\frac{1}{8}$} \, {\rm{tr}} \, w.
\end{equation}

Our calculation of the deviation $\Delta W=\big<W\big>-2$ from the divalent Eu configuration caused by the covalence gives the result:
\begin{equation}
\Delta W=\frac{4}{49}\,\frac{t_0^{\,2}+2\,t_2^{\,2}}{(\Delta E)^2}\,\sum_{J=0}^6\frac{1+2J}{[1+J(J+1)X\,]^2}. \label{ValBed}
\end{equation}

Because of the uncertainty of the precise value of $\Delta W$ [see (\ref{MV}) and (\ref{ValDev})] we carry out the model calculation using the experimental values
\begin{equation}
\Delta W=0.2 \quad \mbox{and} \quad \Delta W=0.15.
\end{equation}

Fixing the value of $\Delta W$ accordingly reduces the number of unknown model parameters from three to two. For convenience, we define
\begin{equation}
t=\sqrt{t_0^{\,2}+2\,t_2^{\,2}}
\end{equation}
as the total hopping amplitude. Because of $t_{-2}=t_2$, $t^2$ is proportional to the total hopping probability of all single-particle Eu states $m$ involved in the hopping. In view of Eq. (\ref{ValBed}), $t$ can be considered a function of $X$ at a given value of $\Delta W$.

In order to ensure that the second-order calculation be consistent, $t/\Delta E$ must be considerably lower than 1. This will be the case for our choices of the model parameters. Higher orders will be suppressed by a factor of $(t/\Delta E)^2$.

\subsection*{Effective Curie susceptibility}

Due to the covalent admixture of the Eu$^{3+}$ configuration the paramagnetic susceptibility of EuRh$_2$P$_2$ can be expected to become anisotropic. According to the space group the susceptibility tensor has tetragonal symmetry. In the high-temperature regime we are considering here the susceptibility is Curie-like. In the limit of small differences between the energies of the Kramers doublets, the paramagnetic susceptibility of the system per space direction $i=x,y,z$ is given by:
\begin{equation}
\chi_{ii}=\mbox{$\frac{1}{kT}$} \, C_{ii},\quad C_{ii}=\mbox{$\frac{1}{8}$}\sum_{MM'}\big|\big<\mbox{$\frac{7}{2}$}M \big| m_i\big| \mbox{$\frac{7}{2}$}M' \big> \big|^{\,2},
\end{equation}
following from the standard formula for the Curie susceptibility of single ions. \cite{abr} The Curie constants $C_{ii}$ of second order depend on the effective magnetic moment  $m_i=\Gamma^\dag M_i\, \Gamma$ of second order in $V$, where $M_i=J_i+S_i$ denotes the untransformed moment. Experimentally, the Curie susceptibility has been measured on polycrystals. \cite{michels,schuette} This corresponds to the spatial average $C=(2\,C_{xx}+C_{zz})/3$, and we define
\begin{equation}
\Delta C_{ii}=C-C_{ii}\quad ({\rm{implying}}\; \Delta C_{xx}\!=\!-\mbox{$\frac{1}{2}$}\Delta C_{zz}).
\end{equation}
\begin{widetext}

{\noindent}The resulting model Curie constants are (in units of $\mu_{\rm{B}}^2$):%\small
\setlength{\arraycolsep}{0mm}
\renewcommand{\arraystretch}{2}
\[
\begin{array}{rlllllll}C=21 -\,\frac{9}{49} \big(\frac{t}{\Delta E}\big)^2&\,  \big[ \;16\; & + \;\frac{1}{1 + 2X} & + \;\frac{5}{1 + 6X} & + \;\frac{14}{1 + 12X} & + \;\frac{30}{1 + 20X} & + \;\frac{55}{1 + 30X} & + \;\frac{91}{1 + 42X}   \\
&  &  +\frac{45}{\left(1 + 2X \right)^{2}} & +\frac{65}{{\left( 1 + 6X \right) }^2} & + \frac{70}{{\left( 1 + 12X \right) }^2} & + \frac{54}{{\left( 1 + 20X \right) }^2} & + \frac{11}{{\left( 1 + 30X \right) }^2} & - \frac{65}{{\left( 1 + 42X \right) }^2} \big],\\
\Delta C_{zz}=\frac{6}{49}\big(\frac{t_0}{\Delta E}\big)^2 &\;\big[  \; 24\; &  + \;\frac{6}{1 + 2X} & + \;\frac{26}{1 + 6X} & + \;\frac{56}{1 + 12X} & + \;\frac{72}{1 + 20X} & + \;\frac{22}{1 + 30X} & - \;\frac{182}{1 + 42X} \\
& & + \frac{63}{\left(1 + 2X \right)^{2}} & + \frac{77}{{\left( 1 + 6X \right) }^2} & + \frac{56}{{\left( 1 + 12X \right) }^2} &  - \frac{77}{{\left( 1 + 30X \right) }^2} & - \frac{143}{{\left( 1 + 42X \right) }^2} & \big].
\end{array}
\]\normalsize\renewcommand{\arraystretch}{1}
\begin{equation} \label{CurKst}  \end{equation}

\end{widetext}
In this calculation we have included (in contrast to an intermediate-valent hybridization) non-trivial contributions to the Curie susceptibility from the localized p$_z$ orbitals:
\begin{equation}
C= C^{\,\rm{Eu}}+C^{\,\rm{p}_z},\quad C^{\,\rm{p}_z}=6\, \Delta W.
\label{Pcontributions}
\end{equation}

As $t / \Delta E$ can be expressed via $X$ [see Eq. (\ref{ValBed})], using Eqs. (\ref{ValBed}) and (\ref{CurKst}) we can fix $t$ and $X$ (and consequently, $\Delta E$) from the experimental values of $\Delta W$ and $C$.
Similarly to the valence, we choose two experimental values of the Curie constants of the polycrystalline samples: \cite{schuette}
\begin{equation}
C=17.2 \quad \mbox{and} \quad C=17.6.
\end{equation}
These values refer to two different samples. The latter value of $C$ is inconsistent with $\Delta W$=0.2 according to the model. Probably, this is not an objection to the model but a further hint at $\Delta W<0.2$ as far as the valence shift is caused by the covalence. We fix three sets of model parameters $X$ (or $\Delta E$) and $t$ according to Table \ref{ParSaetze}. Notice that the perturbation parameter $(t/\Delta E)^2$ is considerably lower than 1, which shows that the low-order calculation is sufficient.
The first parameter set is the least realistic one because $t$ is extremely high.

Table \ref{ParSaetze} shows that for a given value of $\Delta W$, the total hopping amplitude $t$ is very sensitive to the value of $C$. Hence, it is important to take into account the contributions $C^{\,\rm{p}_z}$ from the P$_2$ molecules to the paramagnetic susceptibility, see Eq. (\ref{Pcontributions}).

For convenience we define the relative hopping amplitude concerning $m=0$ single-particle states:
\begin{equation}
\tau=\frac{t_0}{t}.
\end{equation}

After $t$ and $X$ (and $\Delta E$) have been fixed using the
experimental values of $\Delta W$ and $C$, $\tau$ is the only
unknown parameter of the model. Notice that the {\it{anisotropy of
the Curie constant}}, $\Delta C_{zz}$, is proportional to $\tau^2$
[see Eq. (\ref{CurKst})], i.\,e., this anisotropy is solely caused
by the covalent hybridization which involves $m=0$ single-particle
states of the Eu-f shell.

Our model characterizes intervals (upper bounds) for the anisotropy of the Curie constant. $C_{zz}/C_{xx} -1  \; (\geq 0)$ is almost exactly proportional to $\tau^2$. Table \ref{ParSaetze} lists the maximum anisotropy due to:
\begin{equation}
C_{zz}^{\,\rm{max}}=C_{zz}^{\,}|^{\,}_{\tau=1}\,,\quad  C_{xx}^{\,\rm{min}}=C_{xx}^{\,}|^{\,}_{\tau=1}\,.
\end{equation}

\subsection*{Single-ion anisotropy}

The effective Hamiltonian to second order in the perturbation $V$
gives rise to a single-ion anisotropy of the Eu ion:
\begin{equation}
h_{\rm{si}}^{\hspace*{0em}}=\mu \sigma_{z}^2. \label{muEq}
\end{equation}
Here and in the following we normalize every spin operator with
respect to 1 as $\sigma_{\bullet}=2S_{\bullet}/7$. The single-ion
anisotropy parameter $\mu$ is given in Table \ref{mu}. The
anisotropy changes sign for small values of $\tau$. We will come
back to the single-ion anisotropy in the next section when we
discuss various effects on the magnetic ordering temperature,
which are present in EuRh$_2$P$_2$.

\section{The coupling of neighbouring Eu spins}

\subsection*{Nearest neighbours}

The superexchange dynamics of nearest neighbouring Eu spins in EuRh$_2$P$_{\!2}$ is described to leading order by processes of fourth order in the Eu--P$_{\!2}$ hopping within a cluster as shown in Figure 3.

\begin{figure}
\includegraphics[width=7cm]{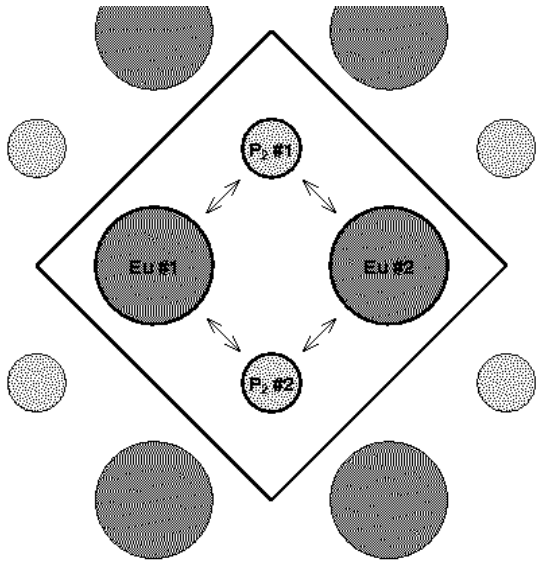}
\label{cluster}
\caption{Quasi two-dimensional EuP$_{\!2}$ plane with a superexchange cluster of nearest neighbouring Eu ions and their common P$_2$ molecular neighbours.}
\end{figure}

In treating the intermediate states of the perturbation series, we use the same Land$\acute{\mbox{e}}$ levels as in section 2 and ignore Coulomb repulsion within the P$_2$ molecules in case of doubly occupied P$_{\!2}$ orbitals. In analogy to Eq. (\ref{perturbation}), the hybridization operator is given by:
\begin{equation}
V=\hspace*{-0.3em}\sum_{i,k=1,2 \atop
\sigma=\uparrow,\downarrow}\hspace*{-0.3em}
\big\{t_0^{\nil}f^{(i)\dag}_{0\sigma}+(-1)^k\,t_2^{\nil}
\big[f^{(i)\dag}_{-2\sigma}+f^{(i)\dag}_{2\sigma}\big]\big\}\,p^{(k)}_\sigma+\mbox{H.\,c.}
\end{equation}
$i$ is a site index for the Eu ions. We calculate the effective superexchange Hamiltonian of fourth order in $V$ which scales with $t^4(\Delta E)^{-3}$ and describes any superexchange in the crystal to leading order. Superexchange processes between nearest neighbouring Eu ions mediated by hopping paths exceeding the cluster we consider are at least of the order $t^6(\Delta E)^{-5}$.

We express the matrix elements of the effective superexchange Hamiltonian in terms of the same quantities as in section 2.
The result is a finite polynomial in terms of spin operators.
To a good approximation the superexchange Hamiltonian turns out as an $xxz$ model:
\begin{equation}
h_{xxz}=j_x\big(\sigma_{1x} \sigma_{2x}+\sigma_{1y} \sigma_{2y}\big)+j_z\,\sigma_{1z} \sigma_{2z}.
\end{equation}

There are additional (multilinear, see below) parts of the
superexchange Hamiltonian, e.\,g., terms $\sigma_{1i}^2
\sigma_{2i}^2$, whereas the tetragonal symmetry of the
superexchange Hamiltonian is exact due to the crystal symmetry.
The hopping processes we consider have got complicated selection
rules. For instance, exchange processes of fourth order in $V$ are
possible where the $z$ component of one Eu spin is changed from
$-3/2$ to $7/2$, i.\,e., these processes have the selection rule
$\Delta S_z=5$. This is why the exchange Hamiltonian is not
bilinear in the spin operators but multilinear. For a bilinear
exchange Hamiltonian the selection rule $\Delta S_z= \pm 1$ is
required. However, the exchange Hamiltonian we obtain is bilinear
to a good approximation. The values of the coefficients of the
neglected part of the Hamiltonian depend on the choice of the
parameter sets: less than $10^{-3}j_x$, $5\times 10^{-2}j_x$, and
$10^{-1}j_x$ for $X=0.00181$, $0.00989$, and $0.0177$,
respectively.

The---antiferromagnetic---coupling constants are listed in terms of {\it{approximate}} numbers in Table \ref{CC1}. The precision of these numbers decreases with increasing value of $X$ as the neglected parts of the Hamiltonian get more important. The lower the values of $X$ and $\tau$, the more isotropic is the $xxz$ coupling. The contribution, which is proportional to $t_0^{\,2}t_2^{\,2}$, is negligible for every parameter set.
Figure 4 
shows the coupling, which has a considerable strength in each case, in terms of the spatial average $j=(2j_x +j_z)/3$.

\begin{figure}[htb]
\includegraphics[width=7cm]{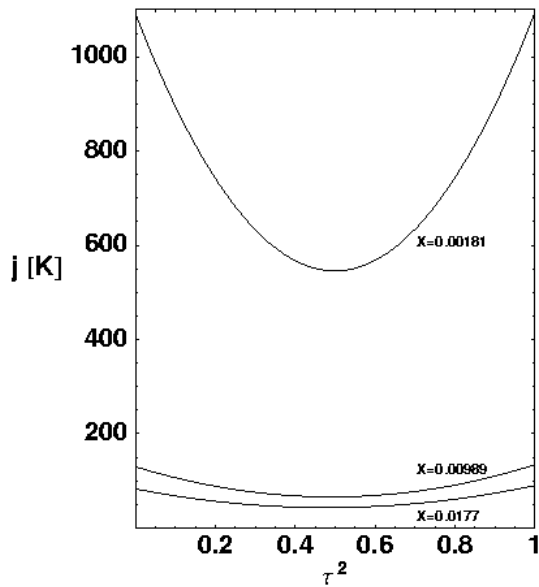}
\label{couplcst}
\caption{Average coupling constant of nearest neighbours.}
\end{figure}

\subsection*{Next-nearest neighbours}

The calculation of the coupling of next-nearest neighbours (these are located within the EuP$_2$ planes) is carried out using almost the same formalism as for nearest neighbours. The only difference is the single P$_{\!2}$ molecule ion involved in this case.

The resulting superexchange Hamiltonian for next-nearest neighbours is denoted by:
\begin{equation}
h'_{xxz}=j'_x\big(\sigma_{1x}^{\nil} \sigma_{2x}^{\nil}+\sigma_{1y}^{\nil} \sigma_{2y}^{\nil}\big)+j'_z\,\sigma_{1z}^{\nil} \sigma_{2z}^{\nil}.
\end{equation}

The calculated coupling constants are listed in Table \ref{ccstnnn}. In contrast to $h_{xxz}$, now the $xx$ part $j'_x$ of the coupling is approximately independent of $\tau$ whereas the Ising part {$j'_z$} is lower than $j_z$ exactly by a factor of four.

\subsection*{Competing effects on the magnetic ordering temperature}

As the value of the model parameter $\tau$, the coupling between
different EuP$_2$ planes and the effect of the indirect exchange
(mediated by delocalized conduction band electrons) on the intraplanar
spin couplings are not known, we cannot
present a quantitative calculation of the magnetic ordering temperature.
However, in the
following we will argue why the considerable value of 50 K for the
ordering temperature is {\em{generic}} in the framework of our
model.

\begin{figure}
\includegraphics[width=7cm]{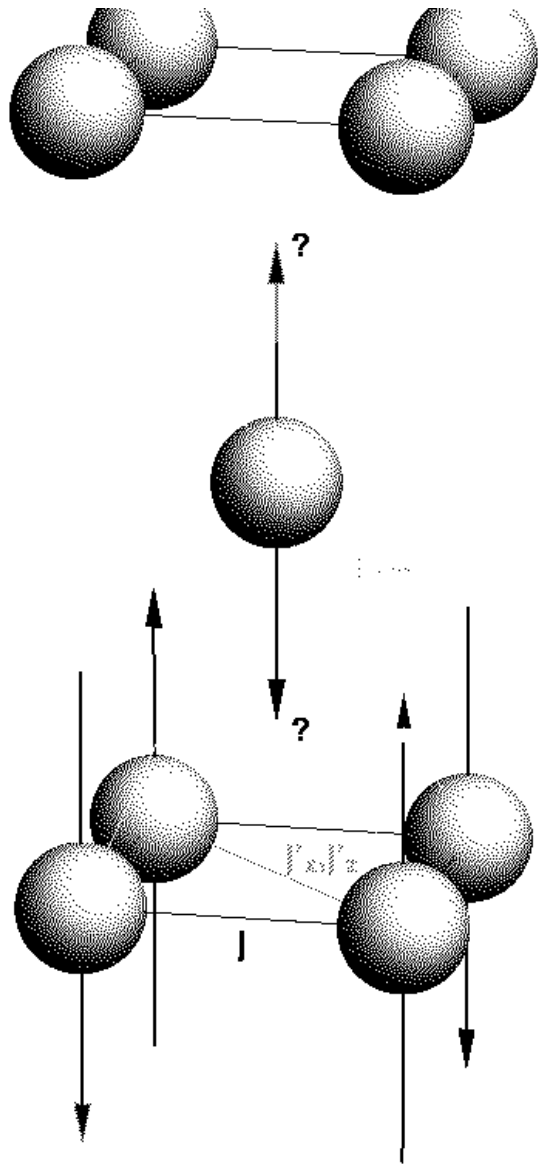}
\label{fr}
\caption{The frustration of the Eu lattice, caused by
antiferromagnetic intraplanar Heisenberg couplings $j$. An example
of four antiferromagnetically ordered spins in a EuP$_{\!2}$ plane
is shown. (i) There is a frustration caused by the
antiferromagnetic next-nearest neighbour couplings $j'_x$ and
$j'_z$ within a EuP$_2$ plane. (i) A nearest neighbour belonging
to a different plane is geometrically frustrated independently of
the sign of $j_{\,\rm{inter}}$.}
\end{figure}

There are competing effects on the magnetic ordering temperature
of EuRh$_2$P$_2$. (i) There is a twofold frustration of the
magnetic interactions caused by the antiferromagnetic next-nearest
neighbour couplings in the EuP$_2$ planes and by the coupling
between different EuP$_2$ planes irrespective of the sign of the
Heisenberg coupling $j_{\,\rm{inter}}$ between Eu ions in
different planes, see Figure
5. The frustration tends to {\em{decrease}} the magnetic ordering
temperature. \cite{toulouse1} (ii) On the other hand, there may be
effects that tend to {\em{enhance}} the ordering temperature.
Except for certain small values of $\tau$, one of these effects is
the single-ion anisotropy (described by the parameter $\mu$, see
Eq. \ref{muEq} and Table \ref{mu}), which may have considerable
strength compared to the superexchange couplings and may be more
relevant than the $xxz$ anisotropies of these couplings. The
ordering temperature is enhanced drastically, i.~e.,
logarithmically by a single-ion anisotropy. \cite{khokh} 
The unknown Heisenberg exchange coupling $j_{\,\rm{inter}}$ or an
unknown dipolar interaction between different EuP$_2$ planes
has---on the mean-field level---no effect on the ordering
temperature as a hypothetically given Ne${\acute{\rm{e}}}$l order
in one plane would not cause a mean field on a Eu site in a
neighbouring plane. However, the unknown anisotropies of the
interplanar exchange and pseudodipolar couplings between
neighbouring planes tend to stabilize the magnetic order. \cite{ahab}

\subsection*{Ordering temperature of an isolated quasi two-dimensional
EuP$_2$ plane}

Though quantitative estimates of various magnetic couplings which may have effects on the magnetic ordering temperature as discussed above are missing, we can present estimates for that temperature according to the coupling in the quasi two-dimensional EuP$_2$ planes. This can serve as the {\it starting point} for a more comprehensive analysis in the future, in particular including the interplanar couplings which are not known to date.

Ref.~\onlinecite{khokh} takes into account the isotropic Heisenberg couplings $j$ of
nearest-neighbouring magnetic ions on a two-dimensional square lattice as well
as the single-ion anisotropies $\mu$ and estimates the magnetic ordering
temperature based on these two parameters for $\mu<<j$:
\begin{equation}
T_{\rm{N}}=\frac{12j}{\big| {\rm ln} |\frac{\mu}{j}| \big|}. \label{TN}
\end{equation}

This Equation gives already considerable ordering temperatures except for very
narrow windows of our unknown model parameter $\tau$, namely
$T_{\rm{N}}> 200$ K
except for $\tau \lesssim 10^{-12}$, $0.13 \lesssim \tau \lesssim 0.15$ and
$0.21 \lesssim \tau \lesssim 0.22$ for the model parameter sets with
$X=0.00181$, $X=0.00989$ and $X=0.0177$, respectively.

Altogether the discussion of the competing effects on the ordering temperature
and of the intraplanar estimate (\ref{TN}) shows that the experimentally
observed ordering
temperature of about 50 K is consistent with our calculations because in this
sense our model avoids the suppression of the ordering temperature
which is implied by intermediate-valent models. \cite{nolting2}

\section{Summary}

We have introduced a systematic interpretation of the Eu valence
shift and the magnetism of EuRh$_2$P$_{\!2}$ due to covalent
bonding which---in contrast to a hypothetically given intermediate
valence---is consistent with experiment. We have presented a model
for the covalence which predicts upper bounds of the anisotropy of
the Curie constants and which characterizes the strength of the Eu
single-ion anisotropies and of the superexchange coupling between
nearest and next-nearest neighbouring Eu ions. Though a quantitative
calculation of the magnetic ordering temperature has not been
possible, we have argued why the experimentally observed ordering
temperature is generic, because for instance the single-ion
anisotropies might have considerable strength.

Measurements of the anisotropy of the Curie constants could fix
the last free parameter $\tau$ of the single-ion anisotropy and
superexchange model and determine the model completely. Following
that, measurements of the magnetic structure and the magnetic
excitations via neutron scattering could make a description
possible which also takes into account indirect exchange between
the Eu ions.
(The reader should be reminded that neutron scattering requires the
particularly expensive isotope Eu-153 because the standard isotope Eu-151
absorbs neutrons too strongly. \cite{holmo})
In this way, there is the chance to understand the
ordering mechanism in EuRh$_2$P$_{\!2}$ comprehensively.

\begin{acknowledgments}

We thank M\,M Abd-Elmeguid and A Aharony for helpful discussions.

\end{acknowledgments}

\setlength{\arraycolsep}{0.2cm}
\renewcommand{\arraystretch}{1.5}
\begin{table}
\caption{Selected sets of model parameters and maximum Curie anisotropy.}
%\begin{indented}
\label{ParSaetze}
\begin{ruledtabular}
\begin{tabular}{llllll}
%\br
$\Delta W$ & $C$ & $X$ & $\Delta E$  & $t$ & $C_{zz}^{\,\rm{max}}/C_{xx}^{\,\rm{min}}$ \\
%\mr
0.2 & 17.2 & 0.00181 & 104687\,K & 24402\,K   & 1.018 \\
0.15 & 17.6 & 0.00989 & $\;$\,19163\,K & $\;$\,4505\,K & 1.065 \\
0.15 & 17.2 & 0.0177 & $\;$\,11083\,K & $\;$\,2915\,K & 1.109 \\
%\br
\end{tabular}
%\end{indented}
\end{ruledtabular}
\end{table}
\renewcommand{\arraystretch}{1}
\setlength{\arraycolsep}{0cm}

\renewcommand{\arraystretch}{1.5}
\setlength{\tabcolsep}{0.3cm}
\begin{table}
\begin{ruledtabular}
\caption{Single-ion anisotropy parameter.}
%\begin{indented}
\label{mu}
\begin{tabular}{ll}
%\br
$X$  & $\;$\,$\mu$\,[K] \\
%\mr
0.00181  &  14 $\tau^2$ \\
0.00989 & --1+51 $\tau^2$ \\
0.0177 & --4+87 $\tau^2$ \\
%\br
\end{tabular}
%\end{indented}
\end{ruledtabular}
\end{table}
\setlength{\arraycolsep}{0cm}

\renewcommand{\arraystretch}{1.5}
\setlength{\tabcolsep}{0.02cm}
\begin{table}
\caption{Anisotropic coupling constants of nearest neighbours.}
%\begin{indented}
\label{CC1}
%\item[]
\begin{ruledtabular}
\begin{tabular}{lll}
%\br
$X$  & $\;$\,$j_x$\,[K] & $\;$\,$j_z$\,[K]  \\
%\mr
0.00181  & 1091\,$\tau^{\,4}$+1091\,$(1-\tau^{\,2})^2$ & 1094\,$\tau^{\,4}$+1091\,$(1-\tau^{\,2})^2$  \\
0.00989 & $\;$\,133\,$\tau^{\,4}$+$\;$\,130\,$(1-\tau^{\,2})^2$ & $\;$\,136\,$\tau^{\,4}$+$\;$\,130\,$(1-\tau^{\,2})^2$  \\
0.0177 & $\;\;\;$89\,$\tau^{\,4}$+$\;\,\;$\,83\,$(1-\tau^{\,2})^2$ & $\;\,\;$\,94\,$\tau^{\,4}$+$\;\,\;$\,83\,$(1-\tau^{\,2})^2$  \\
%\br
\end{tabular}
%\end{indented}
\end{ruledtabular}
\end{table}
\setlength{\arraycolsep}{0cm}

\renewcommand{\arraystretch}{1.5}
\begin{table}
\caption{Coupling constants of next-nearest neighbours.}
%\begin{indented}
\label{ccstnnn}
%\item[]
\begin{ruledtabular}
\begin{tabular}{lll}
%\br
$X$  & $\;$\,$j'_x$\,[K]& $\;\,j'_z$\,[K]\\
%\mr
0.00181 & 273 & 273\,$\tau^{\,4}$+273\,$(1-\tau^{\,2})^2$\\
0.00989 & $\;$\,33 & $\;$\,32\,$\tau^{\,4}$+$\;$\,34\,$(1-\tau^{\,2})^2$\\
0.0177 & $\;$\,22 & $\;$\,21\,$\tau^{\,4}$+$\;$\,23\,$(1-\tau^{\,2})^2$\\
%\br
\end{tabular}
%\end{indented}
\end{ruledtabular}
\end{table}
\renewcommand{\arraystretch}{1}

\end{document}